\documentclass[twocolumn,epjc3]{svjour3}          

\usepackage[T1]{fontenc}
\usepackage[UKenglish]{babel}
\usepackage{graphicx}
\usepackage{mathptmx}      
\usepackage{flushend}
\usepackage[numbers,sort&compress]{natbib}
\usepackage[colorlinks,citecolor=blue,urlcolor=blue,linkcolor=blue]{hyperref}

\journalname{Eur.~Phys.~J.~C}

\usepackage{amsmath,amssymb,bbm}

\newcommand{\vx}{{\vec{x}}}
\newcommand{\vy}{{\vec{y}}}
\newcommand{\vz}{{\vec{z}}}
\newcommand{\vp}{{\vec{p}}}
\newcommand{\vq}{{\vec{q}}}
\newcommand{\uvp}{{\hat{\vec{p}}}}
\newcommand{\uvq}{{\hat{\vec{q}}}}
\newcommand{\vl}{{\vec{\ell}}}
\newcommand{\valpha}{{\vec{\alpha}}}
\newcommand*{\bra}[1]{\langle #1 \rvert}
\newcommand*{\ket}[1]{\lvert #1 \rangle}
\newcommand*{\smallexp}[1]{{{\scriptscriptstyle(\mkern-2mu#1\mkern-2mu)}}}
\newcommand*{\perket}[2]{ \lvert #1 \rangle \mkern-4mu {\vphantom{\big\vert}}^\smallexp{#2}}
\newcommand*{\perbra}[2]{ {\vphantom{\big\vert}}^\smallexp{#2} \mkern-5mu \langle #1 \rvert }
\newcommand*{\perbraket}[4]{{\vphantom{\big\vert}}^\smallexp{#2}\mkern-5mu\langle#1\vert#3\rangle\mkern-4mu{\vphantom{\big\vert}}^\smallexp{#4}}
\DeclareMathOperator{\Det}{Det}	
\DeclareMathOperator{\tr}{tr}	
\renewcommand*{\d}[1][]{\mathop{\mathrm{d}^{#1}}\mkern-4mu} 	
\newcommand*{\I}{{\ensuremath{\mathrm{i}}}}
\newcommand*{\e}{{\ensuremath{\mathrm{e}}}}
\newcommand*{\Nc}{N_\mathrm{c}}
\newcommand*{\be}{\begin{equation}}
\newcommand*{\ee}{\end{equation}}
\newcommand*{\tsqrt}[1]{\sqrt{\smash[b]{#1}}}
\newcommand*{\Eqref}[1]{Eq.~\eqref{#1}}
\newcommand*{\coloneq}{\mathrel{\mathop:}=}
\newcommand*{\eqcolon}{=\mathrel{\mathop:}}
\newcommand*{\abs}[1]{\ensuremath{\lvert#1\rvert}}
\newcommand*{\deltabar}{\delta\mkern-8mu\mathchar'26}
\newcommand*{\dfr}[2][]{{\ifx&#1&\frac{\mathrm{d}#2}{2\pi}\else\frac{\mathrm{d}^{#1}#2}{(2\pi)^{#1}}\fi}}
\newcommand*{\eps}{\varepsilon}
\newcommand*{\dbar}{\text{\dj}}

\begin{document}

\title{Hamiltonian Approach to QCD in Coulomb Gauge: Perturbative Treatment of the Quark Sector}
\author{Davide R. Campagnari\thanksref{e1} \and Hugo Reinhardt\thanksref{e2}}
\thankstext{e1}{e-mail: d.campagnari@uni-tuebingen.de}
\thankstext{e2}{e-mail: hugo.reinhardt@uni-tuebingen.de}
\institute{Institut f\"ur Theoretische Physik, Universit\"at T\"ubingen, Auf der Morgenstelle 14, 72076 T\"ubingen, Germany}
\date{\today}

\maketitle

\begin{abstract}
We study the static gluon and quark propagator of the
Hamiltonian approach to Quantum Chromodynamics in Coulomb gauge in one-loop
Rayleigh--Schr\"odinger perturbation theory. We show that the results agree with the equal-time limit
of the four-dimensional propagators evaluated in the functional integral (Lagrangian) approach.
\end{abstract}


\section{Introduction}

Over the years there has been increased activity in Quantum Chromodynamics (QCD)
in Coulomb gauge, starting with the seminal works of Gribov \cite{Gribov:1977wm} and
Zwanziger \cite{Zwanziger:1998ez}. The use of Coulomb gauge is motivated by the fact
that it is a so-called ``physical gauge''. In fact, in QED Coulomb gauge fixing yields immediately
the gauge invariant (transverse) part of the gauge field, i.e.~physical degrees of freedom.
Although this is not the case in QCD, one still expects that the transverse components of the 
gauge field contain the dominant part of the physical degrees of freedom.\footnote{This is indeed confirmed by variational
calculations within the Hamiltonian approach to Yang--Mills theory in Coulomb gauge, where the longitudinal part of the momentum operator,
i.e.~of the kinetic energy, is completely irrelevant compared to the transverse part \cite{Heffner:2012sx}.}

Coulomb gauge has been mainly used in two approaches to QCD: 
i) in the Dyson--Schwinger equations (DSEs) based on the functional integral formulation
of QCD \cite{Zwanziger:1998ez,Baulieu:1998kx,Watson:2006yq,Watson:2007mz,Watson:2008fb,Popovici:2008ty,Watson:2011kv,Popovici:2010mb}
and 
ii) in a variational approach based on the Hamiltonian formulation
\cite{Schutte:1985sd,Szczepaniak:2001rg,Feuchter:2004mk,Reinhardt:2004mm,%
Epple:2006hv,Epple:2007ut,Campagnari:2010wc,Pak:2011wu,Reinhardt:2011hq,%
Reinhardt:2012qe}.
The general formulation of Yang--Mills theory within the Dyson--Schwinger approach in Coulomb gauge
was set up in Ref.~\cite{Watson:2006yq} and treated in one-loop perturbation theory in
Refs.~\cite{Watson:2007mz}. Thereby the results of covariant gauges were
reproduced.

Since QCD is an asymptotically free theory, one expects its high-energy
behaviour to be dominated by the perturbative results. The understanding
of perturbation theory is therefore necessary for the regularization and renormalization
of non-perturbative approaches. The perturbative treatment of the Yang--Mills sector of QCD
within the Hamiltonian formulation was given in Refs.~\cite{Campagnari:2009km,Campagnari:2009wj}. 
In the present paper we extend the perturbative analysis to the
quark sector of QCD. Some of the results presented below have been already obtained in Ref.~\cite{PhDCampagnari}.
Using the familiar Rayleigh--Schr\"odinger perturbation theory we calculate
the static quark and gluon propagators to one-loop order within the Hamil\-tonian approach
and show that they agree with the equal-time limit of the four-dimensional propagators evaluated in the more traditional
functional integral approach \cite{Popovici:2008ty}. The perturbative results obtained
in the present paper for the static propagators are essential ingredients for the
renormalization of the (non-perturbative) variational approach to the Hamiltonian
formulation of QCD in Coulomb gauge to be presented elsewhere \cite{Campagnari:tbp}.


\section{\label{sec:ham}Perturbative Expansion of the QCD Hamilton Operator}

The Hamilton operator of QCD in Coulomb gauge \cite{Christ:1980ku} reads in $d$ space dimensions
\begin{multline}\label{hamQCD}
H_{\mathrm{QCD}} =
- \frac12 \int \d[d]x \: \mathcal{J}_A^{-1} \frac{\delta}{\delta A_i^a(\vx)} \mathcal{J}_A \frac{\delta}{\delta A_i^a(\vx)} \\
+ \frac14 \int \d[d]x \: B_{i}^a(\vx) \, B_{i}^a(\vx) \\
+\int \d[d]x \: \psi^{m\dag}(\vx) \bigl[-\I \alpha_i \partial_i + \beta \, m\bigr] \psi^m(\vx) \displaybreak[1] \\
- g \int \d[d]x \: \psi^{m\dag}(\vx) \alpha_i A_i^a(\vx) \, t^a_{mn} \psi^n(\vx) \displaybreak[1] \\
+ \frac{g^2}{2} \int \d[d]x \d[d]y \: \mathcal{J}_A^{-1} \rho^a(\vx) \mathcal{J}_A \, F_A^{ab}(\vx,\vy) \, \rho^b(\vy)
\end{multline}
where $\alpha_i$ and $\beta$ are the Dirac matrices, $t^a$ (with $a=1,\dots,\Nc^2-1$) are the
generators of the $\mathfrak{su}(\Nc)$ algebra in the fundamental representation, and
\be\label{fp}
\begin{split}
&\mathcal{J}_A = \Det G_A^{-1} , \\
&[G_A^{ab}(\vx,\vy)]^{-1} = \bigl( - \delta^{ab} \partial^2 - g f^{acb} A_i^c(\vx) \partial_i \bigr) \delta(\vx-\vy)
\end{split}
\ee
is the Faddeev--Popov determinant, with $g$ being the bare coupling and
$f^{abc}$ the structure constants of $\mathfrak{su}(\Nc)$. Furthermore,
$B_{i}^a$ is the non-Abelian magnetic field, and
\be\label{coulkernel}
F_A^{ab}(\vx,\vy) = \int \d[d]z \: G_A^{ac}(\vx,\vz) \, (-\partial^2_\vz) \, G_A^{cb}(\vz,\vy)
\ee
is the so-called Coulomb kernel, which arises from the resolution of Gauss's law in
Coulomb gauge: It describes the Coulomb-like interaction between colour charges, whose
density is given by
\[
\begin{split}
\rho^a(\vx) &= \rho^a_\mathrm{Q}(\vx) + \rho^a_A(\vx) \\
&= \psi^{m\dag}(\vx)\, t^a_{mn}\psi^n(\vx) + f^{abc} A_i^b(\vx) \, \frac{\delta}{\I \, \delta A_i^c(\vx)}
\end{split}
\]
to which both the quarks and the gluons contribute.

The fermion field operator $\psi$ can be expanded in terms of the eigenspinors $u(\vp,s)$,
$v(\vp,s)$ of the free Dirac Hamiltonian
\[
h_0(\vp) = \valpha\cdot\vp + \beta \, m
\]
in the standard way
\be\label{dir2a}
\begin{split}
\psi^m(\vx) &= \int \dbar{p} \: \e^{\I\vp\cdot\vx} \psi^m (\vp) , \\
\psi^m (\vp) &= \frac{1}{\tsqrt{2 E_\vp}} \bigl[ u (\vp,s) \, b^m(\vp,s) + v (-\vp,s) \, d^{m\dag}(-\vp,s) \bigr] ,
\end{split}
\ee
where the index $s = \pm 1$ accounts for the two spin degrees of freedom. Furthermore,
we have introduced the abbreviation
\[
\int \dbar{p} \equiv \int\frac{\d[d]p}{(2\pi)^d} .
\]
The spinors $u(\vp,s)$, $v(\vp,s)$ satisfy the eigenvalue equation
\be\label{eigen1}
\begin{split}
h_0(\vp) \, u(\vp,s) &= E_\vp \, u(\vp,s) , \\
h_0(\vp) \, v(-\vp,s) &= - E_\vp \, v(-\vp,s) ,
\end{split}
\ee
with $E_\vp = \tsqrt{\vp^2 + m^2}$, and are normalized to
\be\label{spinnorm}
\begin{gathered}
u^\dag(\vp,s) \, u(\vp,s') = 2 E_\vp \, \delta_{ss'} = v^\dag(\vp,s) \, v(\vp,s') , \\
u^\dag(\vp,s) \, \beta \, u(\vp,s') = 2 m\, \delta_{ss'} = -v^\dag(\vp,s) \, \beta \, v(\vp,s') ,  \\
u^\dag(\vp,s) \, v(-\vp,s') = 0 . 
\end{gathered}
\ee
The expansion coefficients $b^m(\vp,s)$, $d^{n\dagger}(\vp,s)$ are annihilation and creation
operators  satisfying the usual anti-com\-mu\-ta\-tion relations 
\begin{gather*}
\bigl\{ b^m(\vp,s) , b^{n\dag}(\vq,t) \bigr\} = \delta^{mn} \delta_{st} \, (2 \pi)^3 \delta(\vp-\vq), \\
\bigl\{ d^m(\vp,s) , d^{n\dag}(\vq,t) \bigr\} = \delta^{mn} \delta_{st} \, (2 \pi)^3 \delta(\vp-\vq),
\end{gather*}
which, with the normalization \Eqref{spinnorm}, ensure that the Fermi field in coordinate space has the required 
anticommutation relation
\[
\{ \psi^m(\vx), \psi^{n\dagger} (\vy) \} = \delta^{mn} \, \delta(\vx-\vy) .
\]
For later convenience it is useful to introduce the following orthogonal projectors
\be\label{qptproj1}
\Lambda_{\pm}(\vp) \coloneq \frac{\mathbbm{1}}{2} \pm \frac{h_0(\vp)}{2 E_\vp} ,
\ee
which are (colour diagonal) Dirac matrices satisfying
\begin{gather*}
\Lambda_{+}(\vp) + \Lambda_{-}(\vp) = \mathbbm{1} , \qquad
\Lambda_{+}(\vp) \, \Lambda_{-}(\vp) = 0 , \\
[\Lambda_{\pm}(\vp)]^2 = \Lambda_{\pm}(\vp).
\end{gather*}
Furthermore, from Eqs.~\eqref{eigen1} and \eqref{qptproj1} follows
\[
\begin{aligned}
\Lambda_{+}(\vp) \, u(\vp,s) &= u(\vp,s), \qquad & \Lambda_{+}(\vp) \, v(-\vp,s) &= 0 , \\
\Lambda_{-}(\vp) \, v(-\vp,s) &= v(-\vp,s), \qquad & \Lambda_{-}(\vp) \, u(\vp,s) &= 0 .
\end{aligned}
\]
The projectors $\Lambda_{\pm}$ are related to the Dirac spinors by the following
completeness relations
\be\label{spinsums}
\begin{split}
&\sum_s \frac{u(\vp,s) \otimes u^\dag(\vp,s)}{2 E_\vp} = \Lambda_{+}(\vp), \\
&\sum_s \frac{v(\vp,s) \otimes v^\dag(\vp,s)}{2 E_\vp} = \Lambda_{-}(-\vp).
\end{split}
\ee

The Hamiltonian \Eqref{hamQCD} can be perturbatively expanded in powers of the coupling
constant $g$,
\[
H_\mathrm{QCD} = H_0 + g H_1 + g^2 H_2 + \dots
\]
Since the perturbative treatment of the gluon
sector within the Hamiltonian approach in Coulomb gauge was already given in Ref.~\cite{Campagnari:2009km}
we will focus here on the perturbative treatment of the quark sector.
The ``unperturbed'' Hamiltonian for the quarks is the free Dirac Hamiltonian, i.e.~the third term on
the r.h.s.~of \Eqref{hamQCD}. Using the decomposition \Eqref{dir2a} of the quark field
and the orthogonality relations \eqref{spinnorm}, it acquires the standard form
\[
H_0^\mathrm{Q} = E_0^\mathrm{Q} + \int\dbar{p} \: E_\vp \bigl[ b^{m\dag}(\vp,s) \, b^m(\vp,s) + d^{m\dag}(\vp,s) \, d^m(\vp,s) \bigr] ,
\]
where $E_0^\mathrm{Q}$ is the (negative divergent) zero-point energy. The vacuum state of the free
Dirac theory is annihilated by the operators $b$ and $d$,
\be\label{qpt5a}
b^m(\vp,s)\ket{0}^{}_\mathrm{Q} = 0, \qquad d^m(\vp,s)\ket{0}^{}_\mathrm{Q} = 0 ,
\ee
and their Hermitian conjugate operators $b^\dag$ and $d^\dag$ generate the eigenstates
of $H_0^\mathrm{Q}$, e.g.
\begin{align*}
H_0^\mathrm{Q} \, b^{m\dag}(\vp,s)\ket{0}^{}_\mathrm{Q} & = (E_0^\mathrm{Q}+E_\vp) \, b^{m\dag}(\vp,s)\ket{0}^{}_\mathrm{Q} , \\
H_0^\mathrm{Q} d^{m\dagger} (\vp, s) \ket{0}^{}_\mathrm{Q} &= (E_0^\mathrm{Q} + E_\vp) d^{m\dag} (\vp, s) \ket{0}^{}_\mathrm{Q} \, .
\end{align*}

The gauge field operator can be expanded as
\be\label{boslad}
\begin{split}
A_i^c(\vx) &= \int \dbar{p} \: \e^{\I\vp\cdot\vx} A_i^c(\vp), \\
A_i^c(\vp) &= \frac{1}{\tsqrt{2\abs{\vp}}} \bigl[ a_i^c(\vp) + a_i^{c\dag}(-\vp) \bigr]
\end{split}
\ee
where $a$ and $a^\dag$ are bosonic ladder operators satisfying
\[
\bigl[ a_i^c(\vp) , a_j^{b\dag}(\vq) \bigr] = \delta^{cb} t_{ij}(\vp) (2\pi)^d \delta(\vp-\vq) ,
\]
with
\[
t_{ij}(\vp) = \delta_{ij} - \frac{p_i \, p_j}{\vp^2}
\]
being the transverse projector in momentum space. The unperturbed gluon Hamiltonian becomes
\[
H_0^\mathrm{YM} = E_0^\mathrm{YM} + \int\dbar{p} \: \abs{\vp} a_i^{c\dag}(\vp) a_i^c(\vp)
\]
where $E_0^\mathrm{YM}$ is the irrelevant diverging zero-point energy of the gluons.
The vacuum state of the free Yang--Mills sector is annihilated by the operators $a$
\be\label{G5}
a_i^c(\vp) \ket{0}^{}_\mathrm{YM} = 0
\ee
and the eigenstates of $H_0^\mathrm{YM}$ are generated by $a^\dag$. 

The unperturbed QCD vacuum state is given by the tensor product
\[
\ket{0} = \ket{0}^{}_\mathrm{YM} \otimes \ket{0}^{}_\mathrm{Q} ,
\]
with the quark and gluonic vacuum defined, respectively, by \Eqref{qpt5a} and \Eqref{G5}.
Expectation values of products of field operators obviously factorize in products of
fermionic and gluonic expectation values, e.g.
\[
\bra{0} A \psi^\dag \psi \psi A \psi^\dag \ket{0} =
{}^{}_\mathrm{YM}\bra{0} A A \ket{0}^{}_\mathrm{YM}
\times
{}^{}_\mathrm{Q}\bra{0} \psi^\dag \psi \psi \psi^\dag\ket{0}^{}_\mathrm{Q} \, ,
\]
for which Wick's theorem holds: therefore, in perturbation theory all matrix elements can
be expressed by the free static gluon
\be\label{cont2}
\bra{0} A_i^a(\vp) A_j^b(\vq) \ket{0} = \delta^{ab} \frac{t_{ij}(\vp)}{2\abs{\vp}} (2\pi)^d \delta(\vp+\vq) ,
\ee
and quark Green functions
\be\label{cont}
\begin{split}
\bra{0} \psi_\alpha^m(\vp) \psi_\beta^{n\dag}(\vq) \ket{0} &= \delta^{mn} (2\pi)^d \delta(\vp-\vq) [\Lambda_+(\vp)]_{\alpha\beta} , \\
\bra{0} \psi_\alpha^{m\dag}(\vp) \psi_\beta^n(\vq) \ket{0} &= \delta^{mn} (2\pi)^d \delta(\vp-\vq) [\Lambda_-(\vp)]_{\beta\alpha} ,
\end{split}
\ee
where the spinor indices have been written out explicitly.

The first-order perturbation is given by the minimal coupling term $\psi^\dag A \psi$,
the fourth term on the r.h.s.~of \Eqref{hamQCD}, and reads in momentum space
\be\label{ham1}
H_1= \int \dbar p_1 \dbar p_2 \:
\psi^{k\dag}(\vp_1) \bigl[ - t^a_{kl} \, \alpha_i A_i^a(\vp_1-\vp_2) \bigr] \psi^l(\vp_2) .
\ee
The second-order perturbation arises from the non-Abelian Coulomb interaction, last
term in \Eqref{hamQCD}. Since this operator
comes with a factor of $g^2$, we can replace the Coulomb kernel by its
lowest-order expression, which is given by the negative inverse Laplacian
[see Eqs.~\eqref{fp} and \eqref{coulkernel} with $g=0$], yielding
\begin{multline}\label{ham2}
H_2 = t^a_{mn} \, t^a_{kl} \: \frac12 \int \dbar{p_1} \dbar{p_2} \dbar{p_3} \: \psi^{m\dag}(\vp_1) \, \psi^n(\vp_2) \\
\times \frac{1}{(\vp_1-\vp_2)^2} \: \psi^{k\dag}(\vp_3) \, \psi^l(\vp_1-\vp_2+\vp_3) .
\end{multline}
We have included here only the Coulomb interaction between fermionic charges: the coupling between
the fermionic and gluonic colour charge through the Coulomb kernel does not contribute to
the propagators to one-loop order and will henceforth be discarded.


\section{Perturbative Corrections to the Vacuum State}

In Rayleigh--Schr\"odinger perturbation theory the vacuum state is expanded in a power
series
\be\label{ps}
\ket{0}^{}_\mathrm{QCD} \sim \ket{0} + g \perket{0}{1} + g^2 \perket{0}{2} + \mathcal{O}(g^3).
\ee
and the perturbative corrections to the wave function are chosen to be orthogonal to the
unperturbed state
\[
\langle 0 \perket{0}{n} = 0 , \qquad n\geq 1 .
\]
The first- and second-order corrections to the vacuum state are
\be\label{pwf}
\begin{split}
\perket{0}{1} &= - \sum \frac{\bra{N} H_1 \ket{0}}{E_N-E_0} \ket{N} \\
\perket{0}{2} &= - \sum \frac{\bra{N} H_2 \ket{0} + \bra{N} H_1 \perket{0}{1} }{E_N-E_0} \ket{N} 
\end{split}
\ee
where $\ket{N}$ stands for a generic $N$-particle state\footnote{For bosonic $K$-particle
states a factor $1/K!$ has to be included to avoid multiple counting.} with energy $E_N$.
Furthermore, $E_0=E_0^\mathrm{YM}+E_0^\mathrm{Q}$ is the energy of the perturbative QCD vacuum,
which cancels, however, in the energy denominators.
The series given in \Eqref{ps} is not normalized. The normalized state reads to the
desired order $\mathcal{O}(g^2)$
\be\label{normvac}
\ket{0}^{}_\mathrm{QCD} = {\left( 1 - \frac{g^2}{2} \, \perbraket{0}{1}{0}{1} \right)} \ket{0} + g \perket{0}{1} + g^2 \perket{0}{2} + \mathcal{O}(g^3) .
\ee
It will not be necessary to evaluate $\perbraket{0}{1}{0}{1}$ explicitly: this term
merely cancels disconnected diagrams occurring in the evaluation of the propagators.

With the help of the projectors $\Lambda_\pm$ [\Eqref{qptproj1}] and of the sum rules
\eqref{spinsums} we obtain
\be\label{qpt6}
\begin{split}
b^{m\dag}(\vp,s) \ket{0} \bra{0} b^m(\vp,s) &=
[\Lambda_+(\vp)]_{\alpha\beta} \, \psi^{m\dag}_\alpha(\vp) \ket{0} \bra{0} \psi^m_\beta(\vp) , \\
d^{m\dag}(\vp,s) \ket{0} \bra{0} d^m(\vp,s) &=
[\Lambda_-(\vp)]_{\alpha\beta} \, \psi^m_\beta(\vp) \ket{0} \bra{0} \psi^{m\dag}_\alpha(\vp) .
\end{split}
\ee
Analogously, in view of \Eqref{boslad}
we have
\be\label{qpt6a}
a_i^{c\dag}(\vp) \ket{0} \bra{0} a_i^{c}(\vp)  =2\abs{\vp} A_i^c(-\vp) \ket{0} \bra{0} A_i^c(\vp),
\ee
where we have used $A^\dag(\vp)=A(-\vp)$.
These relations allow us to express the matrix elements occurring in \Eqref{pwf}
in terms of field operators only, for which we can then use Wick's theorem together with
Eqs.~\eqref{cont2} and \eqref{cont}.

The evaluation of the matrix elements in \Eqref{pwf} with the operators $H_1$ and $H_2$
given by Eqs.~\eqref{ham1} and \eqref{ham2} is now straightforward. As an example, we sketch
here the evaluation of the first-order correction. From the form of the first-order
perturbation [\Eqref{ham1}] follows immediately that we need to consider in the sum in
\Eqref{pwf} the states with one gluon, one quark, and one antiquark. With Eqs.~\eqref{qpt6}
and \eqref{qpt6a} we have therefore
\begin{multline*}
\perket{0}{1} = - \int \dbar\ell_1 \, \dbar\ell_2 \, \dbar\ell_3 \:
\frac{\bra{0} A_j^c(\vl_3) \, \psi^{n\dag}_\gamma(\vl_2) \, \psi^m_\beta(\vl_1) \, H_1 \ket{0} }%
     {E_{\vl_1} + E_{\vl_2} + \abs{\vl_3}} \\
\times
[\Lambda_+(\vl_1)]_{\alpha\beta} \, [\Lambda_-(\vl_2)]_{\gamma\delta} \, 2\abs{\vl_3} \,
\psi^{m\dag}_\alpha(\vl_1) \, \psi^n_\delta(\vl_2) \, A_j^c(-\vl_3) \ket{0} .
\end{multline*}
If we now insert the explicit form \Eqref{ham1}
of the first-order perturbation into the above expression, we are led to an expression
containing the matrix element
\[
\bra{0} A_j^c(\vl_3) \, \psi^{n\dag}_\gamma(\vl_2) \, \psi^m_\beta(\vl_1) \, \psi^{k\dag}(\vp_1) A_i^a(\vp_1-\vp_2) \psi^l(\vp_2) \ket{0} .
\]
The fermionic and gluonic expectation values factorize and can be evaluated with the help
of Eqs.~\eqref{cont} and \eqref{cont2}. For the first-order correction
to the QCD vacuum wave functional we finally obtain
\begin{multline}\label{qpv1}
\perket{0}{1} = \int \dbar{p} \, \dbar{q} \: t^a_{mn} \: A_i^a(\vp-\vq) \\
\times \frac{ \psi^{m\dag}(\vp) \, \Lambda_{+}(\vp) \, \alpha_i \, \Lambda_{-}(\vq) \,\psi^n(\vq)}{E_\vp+E_\vq+\abs{\vp-\vq}} \ket{0} .
\end{multline}
Analogously one calculates the higher-order contributions.
The second-order perturbative correction to the vacuum wave functional contains
a fermionic part
\begin{multline}\label{qpv2q}
\perket{0}{2}_{\,qq} = C_F \int \dbar{p} \, \dbar{q} \:
\frac{1}{E_\vp} \: \psi^{m\dag}(\vp) \Lambda_+(\vp) 
\biggl[ -\frac{S_0(\vq)}{2(\vp-\vq)^2} \\
+ \frac{t_{ij}(\vp-\vq)}{2 \abs{\vp-\vq}} \frac{\alpha_i S_0(\vq) \alpha_j}{E_\vp + E_\vq + \abs{\vp-\vq}} \biggr]
\Lambda_{-}(\vp) \psi^m(\vp) \ket{0} ,
\end{multline}
where $C_F=(\Nc^2-1)/(2\Nc)$ is the quadratic Casimir of the fundamental representation,
as well as a term involving gauge field operators
\begin{multline}\label{qpv2a}
\perket{0}{2}_{\,AA} = \frac{1}{4} \int \dbar{p} \, \dbar{q} \:
\frac{\tr\bigl[ \Lambda_{+}(\vp) \, \alpha_i \, \Lambda_{-}(\vq) \, \alpha_j \bigr]}{E_\vp+E_\vq+\abs{\vp-\vq}} \\
\times \frac{1}{\abs{\vp-\vq}} \: A_i^a(\vq-\vp) \, A_j^a(\vp-\vq) \ket{0} .
\end{multline}
In \Eqref{qpv2q}, $S_0$ is the tree-level static quark propagator
\[
S_0(\vp) = \frac12 \bigl[ \Lambda_+(\vp) - \Lambda_-(\vp) \bigr] = \frac{\valpha\cdot\vp+\beta m}{2 E_\vp} .
\]
Other second-order corrections to the wave functional do not contribute to the propagators
at one-loop order and will be henceforth ignored. Equation \eqref{qpv2q} contributes only to the quark propagator,
while \Eqref{qpv2a} yields a quark-loop term to the gluon propagator. The first-order
term \Eqref{qpv1} gives a one-loop correction to both the gluon and the quark propagator.

With the perturbative corrections to the QCD vacuum state given above it is straightforward
to carry out the perturbative expansion of the quark and gluon propagators analogously to the 
perturbative treatment of the Yang-Mills sector given in Ref.~\cite{Campagnari:2009km}.


\section{One-Loop Perturbative Propagators}

\subsection{Gluon Propagator}

The perturbative correction to the gluon propagator
\[
{}^{}_\mathrm{QCD}\bra{0} A_i^a(\vp) \, A_j^b(\vq) \ket{0}^{}_\mathrm{QCD} \eqcolon \delta^{ab} \, (2\pi)^d \delta(\vp+\vq) \, t_{ij}(\vp)  D(\vp)
\]
can be evaluated by inserting \Eqref{normvac} with the corrections given by Eqs.~\eqref{qpv1}--\eqref{qpv2a}
into the above expression.
The evaluation of the resulting matrix elements is
straightforward. To one-loop level the gluon propagator is obtained in the form
\[
D(\vp) = \frac{1}{2\abs{\vp}} \Bigl[ 1 + g^2 D_A^{(1)}(\vp) + g^2 D_\mathrm{Q}^{(1)}(\vp) + \mathcal{O}(g^4) \Bigr] ,
\]
where
\begin{multline*}
D_A^{(1)}(\vp)=
\frac{g^2 \Nc}{(d-1)\vp^2} \int \dbar{q} \:
\frac{1-(\uvp\cdot\uvq)^2}{\abs{\vq} \, \abs{\vp+\vq}}
\frac{\abs{\vp}+\abs{\vq}}{\bigl( \abs{\vp}+\abs{\vq}+\abs{\vp+\vq} \bigr)^2} \\
\times \left[ (d-1)(\vp^2+\vq^2) + \frac{(d-2)\vp^2 \vq^2 + (\vp\cdot\vq)^2}{(\vp+\vq)^2} \right] \\
- \frac{g^2 \Nc}{4(d-1)\vp^2} \int \dbar{q} \: \frac{d-2+(\uvp\cdot\uvq)^2}{\abs{\vq}}
\frac{\vq^2-\vp^2}{(\vp+\vq)^2}
\end{multline*}
represents the one-loop correction arising from the Yang--Mills sector which was
calculated in Refs.~\cite{Campagnari:2009km,Campagnari:2009wj}, while
\begin{multline*}
D_\mathrm{Q}^{(1)}(\vp) = \frac{t_{ij}(\vp)}{2 (d-1) \vp^2}
\int \dbar{q} \: \frac{2\abs{\vp} + E_\vq + E_{\vp+\vq} }{( \abs{\vp} + E_\vq + E_{\vp+\vq} )^2} \\
\times \tr \bigl[ \Lambda_{+}(\vp+\vq) \, \alpha_i \, \Lambda_{-}(\vq) \, \alpha_j \bigr] 
\end{multline*}
is the contribution of the dynamical quarks. The Dirac trace can be explicitly
taken and results in
\begin{multline}\label{qpt7p}
D_\mathrm{Q}^{(1)}(\vp) = \frac{1}{(d-1) \vp^2}
\int \dbar{q} \: \frac{\abs{\vp} + E_\vq }{E_\vq E_{\vp+\vq}} \\
\times \frac{(d-1)\bigl[E_\vq E_{\vp+\vq} + \vq\cdot(\vp+\vq)+m^2\bigr] - 2 q_i t_{ij}(\vp) q_j}{( \abs{\vp} + E_\vq + E_{\vp+\vq} )^2} \, .
\end{multline}
As shown in Refs.~\cite{Campagnari:2009km,Campagnari:2009wj},
the result of the Rayleigh--Schr\"o\-dinger perturbation theory can
be compared with the result \cite{Popovici:2008ty,Watson:2007mz} of the more conventional perturbation
theory in the Lagrangian (functional integral) approach in Coulomb gauge.
The quark-loop contribution to
the gluon form factor evaluated in Ref.~\cite{Popovici:2008ty} reads
\begin{multline}\label{qpt8}
W_\mathrm{Q}^{(1)}(p) = \frac{2}{(d-1) p^2} \int \dfr[d+1]{q} \\
\times \frac{(d-1)\bigl[q_4^2 + q_4 p_4 + \vq\cdot(\vq+\vp) + m^2 \bigr]- 2 q_i t_{ij}(\vp) q_j}{(q^2+m^2)[(p+q)^2+m^2]} \, ,
\end{multline}
where $p^2=p_4^2+\vp^2$ is the squared Euclidean four-momentum. The dressing function
of the equal-time propagator is related to the dressing function \Eqref{qpt8} of the
four-dimensional (energy-dependent) propagator by
\be\label{qpt9}
D_\mathrm{Q}^{(1)}(\vp) = 2 \abs{\vp} \int \dfr{p_4} \frac{W_\mathrm{Q}^{(1)}(p)}{p^2} \, .
\ee
Inserting \Eqref{qpt8} into \Eqref{qpt9} and performing the integrals over $p_4$ and
$q_4$, \Eqref{qpt7p} is indeed recovered. Furthermore,
from the non-renormaliza\-tion of the ghost-gluon vertex the well-known first
coefficient of the QCD $\beta$ function
\begin{gather*}
\beta(g) = \mu^2 \frac{\partial}{\partial\mu^2} g^2(\mu) = \frac{\beta_0}{(4\pi)^2} \, g^4 + \dots, \\
 \beta_0=-\frac{11 N_\mathrm{c} -2N_\mathrm{f}}{3}
\end{gather*}
in presence of quarks is recovered in the present Hamiltonian approach.


\subsection{Quark Propagator}

The quark propagator is defined in the Hamiltonian approach by
\begin{multline}\label{qpt4}
\frac12 \: {}^{}_\mathrm{QCD}\bra{0} \bigl[ \psi_\alpha^m(\vp) , \psi_\beta^{n\dag}(\vq) \bigr] \ket{0}^{}_\mathrm{QCD} \\
\eqcolon S_{\alpha\beta}^{mn}(\vp) \, (2\pi)^d \delta(\vp-\vq) .
\end{multline}
The commutator in \Eqref{qpt4} arises from the equal-time limit of the time-ordering present
in the definition of the time-dependent quark Green function.
The perturbative expansion of the quark propagator can be obtained by inserting
\Eqref{ps} into \Eqref{qpt4}, yielding
\begin{multline}\label{qprop1}
S^{mn}(\vp)\deltabar(\vp-\vq) = \delta^{mn} \, S_0(\vp) \deltabar(\vp-\vq) \\
+ \sum_{k+l>0} g^{k+l} \: \perbra{0}{k} \psi^m(\vp) \, \psi^{n\dag}(\vq) \perket{0}{l} .
\end{multline}
The evaluation
of the matrix elements arising from the insertion of Eqs.~\eqref{qpv1} and \eqref{qpv2q}
into \Eqref{qprop1} is somewhat lengthy but straightforward, and results in
\begin{multline}\label{qprop2}
S(\vp) = S_0(\vp)
\left[ 1 - g^2 C_F \int\dbar{q} \frac{1}{(E_\vp+E_\vq+\abs{\vp+\vq})^2} \frac{d-1}{2\abs{\vp+\vq}} \right] \\
+ g^2 \frac{C_F}{E_\vp} \int\dbar{q} \frac{1}{2 (\vp-\vq)^2}
\left[ \frac12 \: S_0(\vq) - 2 S_0(\vp) \, S_0(\vq) \, S_0(\vp) \right] \\
- g^2 \frac{C_F}{E_\vp} \int\dbar{q} \frac{1}{(E_\vp+E_\vq+\abs{\vp+\vq})^2} \: \frac{t_{ij}(\vp+\vq)}{2\abs{\vp+\vq}} \\
\times \biggl[ (E_\vq+\abs{\vp+\vq}) \frac12 \alpha_i \, S_0(-\vq) \, \alpha_j \\
-2(2E_\vp+E_\vq+\abs{\vp+\vq}) S_0(\vp) \,\alpha_i \, S_0(-\vq) \, \alpha_j \, S_0(\vp) \biggr] .
\end{multline}
If the perturbed propagator is parameterized by
\be\label{qprop4}
S(\vp) = \frac{\mathcal{A}(\vp) \, \valpha\cdot\vp + \beta \, \mathcal{B}(\vp)}{2 E_\vp} 
\ee
from \Eqref{qprop2} we can extract the one-loop expressions for the form factors $\mathcal{A}$
and $\mathcal{B}$ by taking the appropriate traces. This results in
\begin{subequations}\label{qpffs}
\begin{multline}
\mathcal{A}(\vp) = 1 - g^2 \frac{m^2 C_F}{2 \vp^2 E_\vp^2} \int \dfr[d]{q} \: \frac{\vp \cdot(\vp+\vq)}{E_\vq (\vp+\vq)^2} \\
- g^2 \frac{C_F}{2 \vp^2 E_\vp^2} \int \dfr[d]{q} \: \frac{1}{E_\vq \abs{\vp+\vq} (E_\vp+E_\vq+\abs{\vp+\vq})^2} \\
\times \biggl\{ 
\left[ (d-3) \vp\cdot\vq + 2 \: \frac{[\vp\cdot(\vp+\vq)][\vq\cdot(\vp+\vq)]}{(\vp+\vq)^2} \right] \\
\qquad\times\bigl[m^2(E_\vp+E_\vq+\abs{\vp+\vq})-\vp^2 E_\vp\bigr] \\
+(d-1) \vp^2 \bigl[ E_\vp^2 E_\vq + m^2 (2E_\vp+E_\vq+\abs{\vp+\vq}) \bigr]  \biggr\}
\end{multline}
for the form factor of the kinetic term, and
\begin{multline}
\mathcal{B}(\vp) = m + m g^2 \frac{C_F}{2 E_\vp^2} \int \dbar{q} \: \frac{\vp \cdot(\vp+\vq)}{E_\vq (\vp+\vq)^2} \\
+ m g^2 \frac{C_F}{2 E_\vp^2} \int \dbar{q} \: \frac{1}{E_\vq \abs{\vp+\vq} (E_\vp+E_\vq+\abs{\vp+\vq})^2} \\
\times \biggl\{
(2E_\vp+E_\vq+\abs{\vp+\vq}) \\
\qquad\left[ (d-3) \vp\cdot\vq + 2 \: \frac{[\vp\cdot(\vp+\vq)][\vq\cdot(\vp+\vq)]}{(\vp+\vq)^2} \right]\\
+(d-1) \bigl[ (\vp^2-m^2)E_\vp + \vp^2 \abs{\vp+\vq} - m^2 E_\vq \bigr] \biggr\}
\end{multline}
\end{subequations}
for the mass term.

As for the gluon propagator, these form factors can be compared with the results of the
Lagrangian approach. In Ref.~\cite{Popovici:2008ty},
the quark propagator (in Euclidean space) was parameterized in the form\footnote{In the
Hamiltonian approach we work with $\psi^\dag$ rather than $\bar\psi=\psi^\dag \beta$.
The formulae presented here differ from the ones in Ref.~\cite{Popovici:2008ty} by
an overall matrix $\beta$.}
\be\label{qptl1}
S(\vp, p_4) = \frac{p_4 F_t(p) + \valpha\cdot\vp \, F_s(p)+ \beta M(p)}{p_4^2+\vp^2+m^2}
\ee
where the dressing functions $F_t$, $F_s$, and $M$ are functions of $\vp^2$ and $p_4^2$. In this
parameterization the $p_4 p_i$ component has been discarded, since it does not arise at
one-loop level.\footnote{Lattice calculations \cite{Burgio:2012ph}
indicate that this component vanishes.}
The static propagator is obtained from the energy-dependent one [\Eqref{qptl1}] by
integrating out the temporal component $p_4$ of the four-momentum
\be
\label{1336-21}
S(\vp) = \int \dfr{p_4} \: S(\vp, p_4) .
\ee
Since the dressing
function $F_t(p)$ is an even function of $p_4$, this component does not contribute to the equal-time
propagator. Inserting \Eqref{qptl1} into \Eqref{1336-21} we find from \Eqref{qprop4} the identification
\begin{gather*}
\int\dfr{p_4} \frac{F_s(p)}{p_4^2+\vp^2+m^2} = \mathcal{A}(\vp), \\
\int\dfr{p_4} \frac{M(p)}{p_4^2+\vp^2+m^2} = \mathcal{B}(\vp).
\end{gather*}
In fact, inserting here for $F_s (p)$ and $M (p)$ the results found in Ref.~\cite{Popovici:2008ty} and 
performing the integration over $p_4$ we recover for ${\cal{A}} (\vp)$ and $B (\vp)$ the expressions 
given in Eqs.~\eqref{qpffs}.

The renormalization of the quark propagator in the Ray\-leigh--Schr\"odinger perturbation
theory can be worked out in the usual way. For this purpose we write the form factors
(\ref{qpffs}) as
\[
\mathcal{A}(\vp) = 1 + g^2 a_1 , \qquad \mathcal{B}(\vp) = m(1 + g^2 b_1 ) ,
\]
where some regularization scheme has been assumed for the integrals defined by 
Eqs.~\eqref{qpffs}. Inserting these expressions into \Eqref{qprop4} yields
\be\label{qpren2}
S(\vp) = \frac{\valpha\cdot\vp (1 + g^2 a_1) + \beta m (1 + g^2 b_1 )}{2\sqrt{\vp^2+m^2}}.
\ee
Introducing the renormalized mass $m_R$ by
\[
m = Z_m m_R , \qquad Z_m = 1 + g^2 c_1 ,
\]
the propagator \Eqref{qpren2} can be rewritten as
\begin{multline*}
S(\vp) = \left(1-g^2 \frac{c_1 m_R^2}{\vp^2+m^2_R} \right) \\
\times \frac{\valpha\cdot\vp (1 + g^2 a_1) + \beta m_R [1 + g^2 (b_1+c_1) ]}{2\sqrt{\vp^2+\smash[b]{m^2_R}}} .
\end{multline*}
Here the term in the parentheses arises from replacing the bare mass $m$ in
the denominator of \Eqref{qpren2} by the renormalized one $m_R$. Expressing the bare
propagator \Eqref{qpt4} in terms of renormalized quantities by
\[
S(\vp) = Z_2 \: \frac{\valpha\cdot\vp + \beta m_R}{2\sqrt{\vp^2+\smash[b]{m^2_R}}} \, ,
\]
the quark mass and wave function renormalization constants, $Z_m$ and $Z_2$, must be chosen as
\be\label{qpren6}
Z_m = 1+ g^2(a_1 - b_1), \qquad Z_2 = 1 + g^2 \frac{\vp^2 a_1 + m_R^2 b_1}{\vp^2+m_R^2} .
\ee
To obtain the explicit expressions for $a_1$, $b_1$ we use a momentum cut-off in \Eqref{qpffs}.
Alternatively, we can find these quantities by integrating the corresponding results of the
Lagrangian functional integral approach \cite{Popovici:2008ty} over the temporal component $p_4$ of the four-momentum. In this
case, the divergent parts are for $d=3-2\eps$
\be\label{qpren7}
\begin{split}
a_1^\mathrm{div.} &= - \frac{C_F}{(4\pi)^2} \: \frac{\vp^2+4m_R^2}{\vp^2+m_R^2} \: \frac{1}{\eps} \, , \\
b_1^\mathrm{div.} &= \frac{C_F}{(4\pi)^2} \: \frac{2\vp^2-m_R^2}{\vp^2+m_R^2} \: \frac{1}{\eps} \, .
\end{split}
\ee
The coefficients in front of the $\eps$ pole in \Eqref{qpren7} are also found as factors
multiplying $\ln\Lambda^2$ when the integrals in \Eqref{qpffs} are evaluated with a
momentum cut-off $\Lambda$. 

The two form factors $\mathcal{A}$, $\mathcal{B}$ cannot be separately renormalized, since
the coefficients in \Eqref{qpren7} are momentum dependent. However, the quark mass and
wave function renormalization constants [see \Eqref{qpren6}] result in the
momentum-independent quantities
\begin{subequations}
\begin{align}
Z_m &= 1 - \frac{g^2 C_F}{(4\pi)^2} \: \frac{3}{\eps} + \cdots, \label{qpren8a}\\
Z_2 &= 1 - \frac{g^2 C_F}{(4\pi)^2} \: \frac{1}{\eps} + \cdots, 
\end{align}
\end{subequations}
where the renormalization-scheme dependent terms have been discarded. In particular,
\Eqref{qpren8a} is the (at this order perturbatively) gauge-invariant result for the
mass renormalization constant, which agrees with the results for the pole mass from
covariant gauges (see e.g.~Refs.~\cite{Tarrach:1980up}).
Let us also mention that the renormalization procedure carried out above necessitates the 
use of a fermion propagator [\Eqref{qpt4}] defined with the commutator, otherwise
the fermion propagator is not multiplicatively renormalizable.


\section{Conclusions}

In this work we have extended the perturbative analysis of Yang--Mills theory within
the Hamiltonian approach in Coulomb gauge performed in Ref.~\cite{Campagnari:2009km}
to full QCD. The one-loop quark propagator as well as the quark-loop contribution to the
gluon propagator have been calculated. Thereby the equal-time limit of the
time-dependent propagators in the conventional functional integral (Lagrangian) formalism
has been reproduced. Also the one-loop
beta function has been recovered in the present Hamiltonian approach. The perturbative
results obtained in this work are necessary ingredients for the renormalization of the non-perturbative
propagators, which are currently under investigation.


\begin{acknowledgements}
The authors would like to thank P.~Watson for valuable discussions.
This work was supported by the Deutsche For\-schungs\-gemeinschaft (DFG) under contract
No.~Re856/6-3, and by the BMBF under contract No.~06TU7199.
\end{acknowledgements}


\end{document}